# Investigating Information Security Risks of Mobile Device Use within Organizations

**William Bradley Glisson**
School of Humanities
University of Glasgow
Brad.Glisson@glasgow.ac.uk

**Tim Storer**
School of Computing Science
University of Glasgow
Timothy.Storer@glasgow.ac.uk

**ABSTRACT**

Mobile devices, such as phones, tablets and laptops, expose businesses and governments to a multitude of information security risks. While Information Systems research has focused on the security and privacy aspects from the end-user perspective regarding mobile devices, very little research has been conducted within corporate environments. In this work, thirty-two mobile devices were returned by employees in a global Fortune 500 company. In the empirical analysis, a number of significant security risks were uncovered which may have led to leakage of valuable intellectual property or exposed the organization to future legal conflicts. The research contribution is an initial empirical report highlighting examples of corporate policy breaches by users along with providing a foundation for future research on the security risks of the pervasive presence of mobile devices in corporate environments.

**INTRODUCTION**

The mobile device, particularly the smartphone, has become ubiquitous in the information rich corporate environment. Annual growth of smartphone use in the corporate environment is reported to be 22% for 2011 (TNS UK Ltd 2011). Equipped with a mobile device, employees can access the full range of corporate information and services needed to do their job, including emails, business plans, and sales data or customer profiles without the need to ever enter the office. However, the very ubiquity of information availability creates the potential for inappropriate information leakage, should a mobile device be lost, accessed or stolen.

The diverse range of applications installed by default on mobile devices, and the ability to access corporate information from both personal and employer issued devices means that the distinction between personal and professional uses of a device is becoming blurred. Both of these factors are conducive to security risks within an organization and are of growing concern for IT security and corporate compliance departments. To complicate matters, a survey by Goode (2010) Intelligence, in 2009, indicates that just under half of their respondents did not have a specific security policy for mobile phones. This is potentially a problem since human error is commonly recognized as a source of data loss in publicized incidents (CISCO 2008; Poynter 2008; WordPress.com 2009). In addition, the Ponemon Institute (2011) reported survey results indicating that nearly two-thirds of users store a "significant" amount of personal data on a mobile device. The report also notes that 84% of respondents stated that they use the same device for both business and personal applications. The result, as IBM (2011) noted, is that mobile devices present complex security risks and management issues to organizations. The issues that these devices present to organizations potentially have serious financial implications for businesses.

This research presents an exploratory empirical case study that broadly investigates information privacy and security from the perspective of a large organization. The research examined thirty-two mobile devices issued in a global Fortune 500 organization, which had been subsequently returned to the organization by the employees. The research contribution is an initial empirical report highlighting examples of corporate policy breaches along with providing the foundation for future research on the security risks of mobile devices in corporate environments.

The structure of the paper is as follows. Section two discusses relevant corporate security policies. Section three presents the methodology, project details and artifacts. Section four presents an analysis of the results. Section five discusses related work concerning the empirical evaluation of data found on mobile devices. Section six draws conclusions and presents ideas for future work.

**CORPORATE POLICY**

The organization has developed a range of policies covering topics including information exchange, electronic communication, usage and salvaging policies. The *Information Exchange Policy* describes appropriate communication within the organization and to external contractors, customers or regulators. Relevant aspects of the policy decreed that all confidential or sensitive data must be encrypted; mobile devices may be monitored for inappropriate use; the identity of recipients of sensitive or confidential data must be verified; and unauthorized software downloads are not permitted.

The *Electronic Communication Policy* includes a discussion of acceptable use of corporate issued Information and Communication Technologies (ICT). In particular, the policy requires that ICTs should be used to undertake authorized business activities and that personal use should be incidental, limited and consistent with good business practices. In addition, the policy requires that personal use should not interfere with, impact upon or interrupt the efficient, lawful and ethical operation of the business. General guidance offered by the policy is to be polite and use appropriate language, do not embarrass the company, and ensure that the recipients of communications are the intended receivers. The document defines a multitude of activities that are unacceptable. Some of these inappropriate activities included the unauthorized possession or distribution of confidential data; excessive personal use during core working hours; transmitting, accessing, copying or storing excessive non-business related data; accessing unauthorized Internet services; and requesting an employee to divulge their user id and password.

The *Usage Policy* states that mobile devices are tools to be allocated based on need and are not entitlements. The usage policy covers eligibility, the acquisition process and general device restrictions, i.e., restrictions on the number of multiple devices with the same capabilities. The usage policy indicates that issued devices are restricted to business activities unless there is no alternative. It goes on to indicate that personal calls should be minimized and that the company should be reimbursed for the cost of these calls. Premium rate numbers are not allowed unless directly relevant to the business. The policy states that devices purchased for group use is permissible. Device reallocation is encouraged to avoid or reduce procurement costs. The policy does indicate that employees leaving the organization must return Subscriber Identity Module (SIM) cards to their appropriate line manager. It also indicates that misuse of the mobile phones could result in disciplinary action.

A separate *Salvaging Policy* details the actions that should be taken within the organization for phones that are either no longer needed or are not working properly. The policy indicates that it is the responsibility of the user and/or the manager. Actions to be taken included the removal of the SIM card from the device and the erasure of the internal memory of the device so that all telephone numbers, data, pictures and general contact information is removed prior to recycling. The policy has specific instructions for Motorola and Nokia device erasure. The policy provides detailed information on whom to contact within the company and outside vendors used for recycling devices.

**METHODOLOGY**

This research is an exploratory case study of security behavior and information privacy on mobile devices in a specific organization that considers both official policy and residual information artifacts. The identity of the organization is being withheld to ensure anonymity. Preserving anonymity creates an environment where all parties are comfortable presenting commercially sensitive data. The case study is an exploratory analysis of a real-world context comparing evidence of actual action with stated corporate policy through assimilation into the organization for a period of several months. Exploratory case studies help researchers to understand problems in real-world contexts along with identifying future areas of research (Oates 2006). Based on growing industry dependency on mobile devices, it was hypothesized that mobile devices are not compliant with organizational security polices and they are putting organizations at risk. This hypothesis raises several questions for investigation:

- How much corporate information is really at risk on mobile devices?
- What type of data is being stored on corporate mobile devices? To what extent are mobile devices being used for personal, as opposed to corporate, activities?
- To what extent are corporate security policies followed by mobile device users? Do current security policies provide adequate protection for organizations?

To address these questions, the data contents of 32 decommissioned mobile devices that had been issued within the organization were analyzed. The resulting data was then compared with the existing corporate policies to determine if data leakage is potentially an issue within the organization.

*Project Details*

The acquisition method used in this case study is derived from a larger case study examining resold mobile phones purchased from secondary markets (Glisson, Storer, Mayall, Moug and Grispos 2011). For this work, only a single mobile forensic toolkit, XRY Complete Forensic Examiners Kit, was used to acquire data from the devices. This was a pragmatic decision based on the resources available for this case study. A thorough description of the methodology implemented in this research is provided to ease comparison with results from future experiments and to identify explicit limitations. The investigation was divided into three stages which included acquisition of the mobile phones, processing mobile phones, and artifact analysis.

*Acquisition of Devices*

All of the mobile devices were provided by one organization. The devices consisted of Nokia and Motorola models. It did not include Global Positioning System devices, tablets, laptops, or any other devices with an integrated Global System for Mobile Communications capabilities. All of the devices had been issued to employees in the organization and returned for replacement or up-grade.

It is noted that the acquisition method may have created inherent biases in the sample of mobile devices retrieved. The decision to restrict data collection to the phones issued to the employees does not catch corporate data that is stored on personal phones. Restricting the data collection to issued devices may have excluded popular brands or models issued in other organizations. The devices examined did not represent up-to-date technology issued in the organization because the devices had been returned after a period of use. It is interesting that the organization re-issues relatively low-end phones where possible and does not issue smartphones to all employees primarily for financial motivations. Data capture was restricted to a single mobile forensic toolkit.

*Mobile Phone Processing*

Each mobile phone was processed using the mobile forensic toolkit, XRY Complete Forensic Examiners Kit Version 5.5 (XFEK), in a physically secure office. Two laptops were purchased for the project. One laptop was configured with the XFEK toolkit. The other laptop was used to aggregate and analyze the data in a Microsoft Access database. For each device, the XFEK toolkit supports some or all of what are referred to as logical, physical decoded and flash data acquisitions. A logical acquisition recovers logical objects (e.g. files) from a device file system. The XFEK software requests information from the Operating System (OS) and the OS returns active content. A physical acquisition recovers the raw data stored in persistent memory on the device, including deleted information. A decoded physical acquisition reconstructs data to provide more meaningful information. The flash method copies all data from the source system. As Breeuwsma, et al, (2007) note, there is no standard approach to retrieving this type of data.

The end result is that several different perspectives of each processed device were produced in the case study. The detailed process for extracting and organizing the data was as follows:

1. An initial inspection of the appearance of the device was recorded in the database in order to document any visible damage.
2. A picture was taken of the mobile phone from the front and the back of the mobile phone. The back of the mobile phone was captured so that the International Mobile Equipment Identity was visible in the image.
3. A physical acquisition of the phones was attempted with XFEK.
4. A logical acquisition of the phone was attempted with XRY. If the logical extraction of the phone was not successful, then a generic SIM card was introduced into the handset and both extractions were tried again. A majority of the extractions required a generic SIM to be present for the software to work.
5. A manual inspection of the initial data was conducted after each extraction using the XFEK data reader. This step was taken as a precaution against the introduction of illicit data (such as illegal images) into the main data set. If illegal images had been discovered on a device at this stage, the hard disk in the laptop could be removed and provided to the appropriate authorities. A spare hard disk was also configured with XFEK ready for use should the need arise: thankfully this was not necessary. Once checked, the raw extraction results were backed up to an external hard disk and then the next device acquisition took place.
6. The data produced by XFEK was exported to a Microsoft Excel spreadsheet via the conversion capabilities provided by the XFEK data reader.

7. The spreadsheet data was imported into individual tables in a Microsoft Access 2010 database. The data captured from the different types of acquisitions were stored in dedicated tables in a database to maintain consistency with the data classification being produced by XFEK.
8. A backup was made daily and stored on an external USB hard-disk.

The procedure for the investigation intended to capture as many artifacts from the individual mobile devices as possible while attempting to minimize potential data loss.

*Artifact Analysis*

Thirty Nokia's were examined along with two Motorola mobile phones. The manufacturer and model break down for the 32 phones discussed in this area of the case study is available in Table 1 - Manufacture and Model.

| Manufacturer | Model | CNT |
|---|---|---|
| Motorola | RAZR V3i | 1 |
| Motorola | V550 | 1 |
| Nokia | 3109c | 8 |
| Nokia | 6021 | 20 |
| Nokia | 6030 | 1 |
| Nokia | 6300 | 1 |

**Table 1. Manufacture and Model**

Over seven thousand artifacts were returned from logical and physical decoded extractions. The investigation revealed:

- A password was transmitted in clear SMS text.
- Inappropriate conversations in SMS text messages.
- Evidence of unencrypted files being transmitted out of the organization contained in an SMS message: "I emailed you the file with the info for Recipient Name and a couple of questions. Can you have a look? Also I just noticed that the email wasn't encrypted when I press send. I will be more careful"
- A Virtual Private Network (VPN) pass-code saved in a contacts entry with the name of the contact given as 'Vpn'
- Inappropriate pictures and videos.

None of the Nokia or the Motorola phones had password protection or encryption enabled on the devices. The physical decoded extraction of the devices produced the largest number of artifacts from this dataset. An overview of the artifacts is available in Table 2 – Total Artifacts.

|  | **Logical** | **Physical Decoded** | **Flash** |
|---|---|---|---|
| Audio Files | 847 | 365 | 915 |
| Calendar Files | 92 | 93 |  |
| Calls | 833 | 2,901 | 40 |
| Contact | 1,029 | 870 |  |
| Documents | 69 | 13 | 50 |
| Files | 1,788 | 2,242 | 2,731 |
| MMS | 14 |  |  |
| Notes | 5 |  |  |
| Pictures | 1,101 | 552 | 3,349 |
| SMS | 1,795 | 752 | 161 |
| Task | 7 | 3 |  |
| Videos |  |  | 10 |
| Total Artifacts | 7,580 | 7,791 | 7,256 |

**Table 2. Total Artifacts**

The results displayed in Table 3 – Deleted Data indicate that the physical extraction performed by the Flash procedure retrieved more deleted data than the procedure used in the Physical Decoded extraction. In areas where XFEK did not find deleted data, it appears to not have been allocated a column in the data output.

One e-mail password was asked for by what appears to be a fellow employee. The email password was provided in the responding text message. That text message was also sent with another message telling the receiver that the password is all lowercase. Further, the conversation content acknowledges that an employee of the organization lost a laptop with the explanation that it was possibly left on a train.

Potentially inappropriate conversations ranged from discussions about individual user's parents to sexual references. For example, there were 72 references to Mom and Dad in the logical extractions and 9 in the physical extraction. There were five text messages with inappropriate explicit sexual references. Only two of these messages were marked as deleted. In the Physical Flash pictures table in the database, two phones out of the collection contained images labeled with dates.

|  | **Physical Decoded** | **Deleted Physical Decoded** | **Flash** | **Deleted Flash** |
|---|---|---|---|---|
| Audio Files | 365 | No Column | 915 | 520 |
| Calendar Files | 93 | 3 | - | - |
| Calls | 2,901 | 512 | 40 | No Column |
| Contact | 870 | 159 | - | - |
| Docs | 13 | No Column | 50 | 0 |
| Files | 2,242 | 449 | 2,731 | 2,131 |
| Pictures | 552 | No Column | 3,349 | 2,243 |
| SMS | 752 | 206 | 161 | 137 |
| Task | 3 | No Column | - | - |
| Videos | - | - | 10 | 3 |
| Total Artifacts | 7,791 | 1,329 | 7,256 | 5,034 |

**Table 3. Deleted Data**

One phone contained 45 Joint Photographic Group (JPGs) that were labeled with a date. Out of the 45 JPGs that were returned in the query, five had been deleted. A visual examination of the data revealed that several of the images were of a baby, females and a male. Two of these pictures included a female breast feeding the baby. The other device that contained a JPG, named with a date, appears to be a very blurry image of a woman's breast. The image had been deleted on the device. XFEK identified 10 items in the video table. All of these items originated from the same device. Out of these items, three contained the same MD5 hash. Of the remaining seven videos, only three could be viewed and one was of a strip tease.

**RESULTS ANALYSIS**

The results from a relatively small sample size indicate that there are issues that need to be resolved in the organization regarding security policy violations. The fact that the logical acquisition method retrieved over a thousand contacts is, in itself, worrying. This is not only a security violation from the perspective that employees are supposed to ensure that they have removed all information from the device before returning it to the organization as stated in the Salvaging Policy, but it creates the risk of employees being targeted for the data on their mobile device. For example, if an employee is known to work in a certain section of the organization, they could be targeted to acquire contact information on senior management. This contact information can then be used in social engineering attacks.

The exchange of an email password via an SMS message is a clear breach of the Electronic Communications Policy summarized in the Corporate Policy section. To complicate matters, the reason the individual was requesting the information was due to a lost laptop. This highlights the need for all data on mobile devices to be encrypted as is required in the Information Exchange Policy. The references to inappropriate conversations and to language also violate the Information Exchange Policy. The strip tease video and the personal pictures, of what are presumably family members and friends of the employees, also violates both the Information Exchange Policy and the Electronic Communications Policy. While these violations are not especially serious security concerns, they could be of potential embarrassment for any organization. The references to 'Mum', 'Dad' and 'Wife' indicate that the mobile devices are, potentially, not being used in compliance with policies that require personal usage be kept to a minimum. It also raises the question of how much the personal usage is costing the organization. Are employees reimbursing the organization for these text messages?

The initial results of this case study generated a review of the organization's existing security polices in reference to the mobile devices that will be available to employees in the future. It has also initiated a discussion around the issue of mobile devices that are used by multiple users. A common example is a mobile device that is used by a support team and used by the employee who is on-call for a particular period of time and then gives the mobile device to the next person on-call. When there is a security policy violation, how does the organization enforce an acceptable use policy for this group of users? In this scenario, how does the organization ensure that they have identified the appropriate perpetrator?

The data in this case study examined phones that were issued or approved for and purchased by the organization. With the continued blurring of the lines between work related and personal activities, it is only a matter of time before employees start inquiring about the use of personal mobile devices in their organizations. The security issues that this introduces include:

- Who owns the data once it has been transferred to a non-corporate device?
- Does an organization have the right to monitor an employee's activities on a personal device?
- If an investigation needs to take place, does the organization have the legal right to search a privately owned device?
- How does the organization mandate the level of security maintained on the device?
- How does the company minimize data leakage when employees leave the organization?

Prior to this case study the organization did not have a policy explicitly detailing how to handle a mobile phone or internal general device investigation. The results of this case study contributed to the content of a series of new investigation policies, standards and guidelines. The new documentation defines the stakeholders and identifies their roles in the investigation process. They address the broad topics of acquisition of the evidence, establishing and maintaining the integrity of the chain of custody, the storage of the source hardware and data, storage of the target data and reports, along with decommissioning of all hardware and data at the conclusion of the investigation so that it is in compliance with retention legislation.

**RELATED WORK**

An investigation into mobile computing within the Information Systems (IS) discipline indicates that there are relatively few papers on the security of mobile devices, compared with other areas of research in information systems (Ladd, Datta, Sarker and Yu 2010). It also indicates that there is a lack of research in the area of "Mobile Cases and Applications" (Ladd et al. 2010) which specifically examines organizational and societal change and the impact when mobile devices are deployed in organizations. The survey demonstrates that security of mobile devices in corporate environments is an issue which needs to be considered as part of an overall business model.

Business models are continuously evolving with the on-going integration of mobile devices into the work environment. Previous research indicates that the boundaries between work and non-work related activities are continuing to blur (Hislop and Axtell 2011). It has been suggested that business models are changing due to the continued proliferation of mobile devices (Lindgren, Taran and Saughaug 2011). They argue that businesses will need to consider models that address the physical, the digital and the virtual dimensions of the organization. However, it does not discuss the security risk that this evolution introduces to organizations.

Several broad security issues have been highlighted with the introduction of cloud computing as a way to enhance mobile computing capabilities (Kumar and Lu 2010). However, these issues are applicable to all types of mobile users and do not address issues that are specific to small scale mobile device users. In particular, anti-virus software is becoming an increasingly important issue for these devices. It has been noted that anti-virus software is becoming more prevalent for mobile devices (Thing, Subramaniam, Tsai and Chua 2011). They also developed a tool that is a loadable kernel module, monitoring Operating System (OS) calls on an Android operating system (Thing et al. 2011). When the OS makes calls to services to access sensitive data, this activity is detected and the tool records the system calls and presents an alert on the device interface. The papers discussed do not consider critical issues related to mobile devices from a digital forensics perspective or how these issues relate to the corporate environment. Much of the existing work does not utilize empirical data from industry to support their analyses. These investigations, naturally, have an impact on information privacy.

Even though it has been noted that privacy research dates back to the 1960s and 1970s (Smith and Milberg 1996), there is still, a great debate on exactly how to define privacy (Smith, Dinev and Xu 2011). Privacy has been defined as "the interest that individuals have in sustaining a personal space, free from interference by other people and organizations" (Clarke 1999). He goes on to define four privacy elements which included the person, personal behavior, personal communication and personal data. The idea has been put forth that the digitization of communications in today's societies encourages the merger of communication privacy and data privacy into what can be termed information privacy (Belanger and Crossler 2011). Even though there are several definitions of information privacy, these definitions typically entail personal control over the secondary use of personal information (Belanger and Crossler 2011; Belanger, Hiller and Smith 2002; Stone, Gueutal, Gardner and McClure 1983). It is interesting that the concept of privacy in research is taken largely from that of an individual.

Research indicates that many researchers focus on the apprehensions that individuals have regarding the privacy practices of organizations (Belanger and Crossler 2011). However, they do not talk about researching privacy from the perspective of an organization. Privacy research can be classified on levels of analysis which have been defined as societal, organizational group and individual (Smith et al. 2011). However, a large portion of the research is from the view point of securing personal information. While specific research highlights that mobile devices present technical opportunities for security breaches and malicious activities along with identifying investigative challenges (Grispos, Glisson and Storer 2013), there is minimum research that focuses on organizational privacy on mobile devices.

## CONCLUSION AND FUTURE WORK

The increase in mobile subscriptions coupled with the continued ubiquitous integration of mobile devices as a prominent means of communications in technologically advanced societies creates environments conducive to security vulnerabilities. The initial study indicates that mobile phones are putting organizations at risk. The amount of corporate information that is really at risk on mobile devices is potentially substantial. The number of artifacts recovered from logical and physical extractions, as noted in Tables 2 and 3, gives an indication of the scale of this problem. It also indicates that the types of data stored on corporate mobile devices include corporate and personal information. The case study reveals that corporate security policies, followed by mobile device users, are clearly being violated. The initial analysis indicates that current security policies are in place; however, the challenge appears to be the enforcement of policies within this organization.

Future work must address how organizations can minimize risk when utilizing mobile technologies. Short-term solutions include security policy revisions to improve coverage of new technologies and services, as well as to encourage compliance. Potential difficulties arise from cultural resistance to new policies and technical implementations in different environments. These issues need to be investigated from both technical and socio-technical viewpoints.

One solution to these compliance challenges may be the implementation of real-time corporate device monitoring software. This software could have functionality in the form of notifications when inappropriate language is being sent or received via SMS, MMS or emails. Other areas of future research include exploratory research into the development and implementation of mobile phone anti-forensics software, empirically researching the proliferation of viruses on corporate mobiles and detecting the use of public and private clouds on mobile devices.

## ACKNOWLEDGMENTS

The authors would like to thank the Fortune 500 Organization for their support and feedback in this research. Thanks are also expressed to Dr Leif Azzopardi for reviewing this paper.